\documentclass[11pt,a4paper]{article}

\usepackage[T1]{fontenc}
\usepackage{lmodern}
\usepackage{amsmath}

\usepackage[numbers,sort&compress]{natbib}

\usepackage[utf8]{inputenc}
\usepackage{geometry}
\usepackage{setspace}

\usepackage{float}
\usepackage{graphicx}
\usepackage{booktabs}
\usepackage{soul}
\usepackage{textcomp}
\usepackage[hidelinks]{hyperref}
\usepackage{amssymb}
\usepackage{newunicodechar}
\usepackage{siunitx}

\usepackage{authblk}

\newunicodechar{⁻}{\textsuperscript{-}}
\title{Ultraviolet Raman spectroscopy using multimode aluminum oxide waveguides}

\author[1]{Roman Zakoldaev\textsuperscript{*}}
\author[1]{Nikhil Jayakumar}
\author[1]{Elie Vennat-Bresson}
\author[1]{Jean-Claude Tinguely}
\author[2]{Firehun Tsige Dullo\textsuperscript{\textdagger}}
\author[1]{Balpreet Singh Ahluwalia\textsuperscript{\textdaggerdbl}}

\affil[1]{Department of Physics and Technology, UiT The Arctic University of Norway, Klokkargårdsbakken 35, NO-9037 Tromsø, Norway}
\affil[2]{SINTEF Digital Dept. of Microsystems and Nanotechnology, Gaustadallen 23C, 0373 Oslo, Norway}

\date{%
\textsuperscript{*}Corresponding author:
\href{mailto:roman.zakoldaev@uit.no}{roman.zakoldaev@uit.no}\\
\textsuperscript{\textdagger}Email:
\href{mailto:firehun.t.dullo@sintef.no}{firehun.t.dullo@sintef.no}\\
\textsuperscript{\textdaggerdbl}Email:
\href{mailto:balpreet.singh.ahluwalia@uit.no}{balpreet.singh.ahluwalia@uit.no}
}

\begin{document}

\maketitle

\begin{abstract}
Waveguide-based Raman spectroscopy offers a route toward compact and integrated photonic sensors. While most waveguide Raman platforms operate in the near-infrared, extending excitation toward the visible and ultraviolet is attractive because shorter excitation wavelengths can provide substantially stronger Raman scattering and access to resonance-enhanced molecular responses. This transition, however, places greater demands on the photonic material because optical loss, absorption and waveguide-generated spectral background. Here, we investigate aluminum oxide (Al$_2$O$_3$) waveguides as a low-background platform for Raman spectroscopy at 532 nm and 325 nm using a back-collected excitation and detection geometry. At 532~nm, after normalization to the incident excitation power--time product, the measured Al$_2$O$_3$ background was more than two orders of magnitude lower than that of Si$_3$N$_4$. Under 325 nm excitation, the waveguides maintained a weak and reproducible spectral background across the investigated widths. Using immersion oil as a model analyte deposited over the waveguide surface, we demonstrate that analyte-specific Raman fingerprints can be recovered above the waveguide background at both wavelengths. Signal-to-background ratio and spectral contrast were used to quantify analyte visibility relative to the waveguide background, and both metrics consistently confirmed clear detection of the analyte bands. These results demonstrate that Al$_2$O$_3$ combines low measured spectral background with operation from the visible into the ultraviolet, providing a promising route toward integrated Raman spectroscopy at shorter excitation wavelengths.
\end{abstract}


\section{Introduction}

Optical waveguides provide an attractive platform for Raman sensing by integrating multiple functions such as excitation, light-analyte interaction, spectral filtering, and signal collection within a photonic circuits. Such a waveguide platform offers a pathway towards compact, scalable, and potentially multiplexed Raman sensing platforms. In waveguide-enhanced Raman spectroscopy (WERS), the evanescent field associated with the guided excitation mode extends beyond the waveguide surface and interact with analytes in the vicinity of the waveguide interface over millimeter- to centimeter-scale propagation distances, while a fraction of the generated Raman emission is recaptured into guided modes for collection \cite{dhakal2014evanescent,dhakal2015efficiency,ettabib2024waveguide}. Although the interaction is confined to the region sampled by the evanescent field, the extended propagation length increases the effective number of analyte molecules contributing to the detected signal and can therefore compensate for the intrinsically weak Raman scattering process. This approach has been demonstrated for organic compounds \cite{dhakal2014evanescent}, biological submonolayers \cite{dhakal2016biological}, and trace vapors \cite{tyndall2018trace}. Its performance reflects a trade-off between modal overlap with the analyte and propagation loss. 

Waveguide Raman sensing has been demonstrated extensively using silicon nitride (Si$_3$N$_4$) waveguides \cite{dhakal2014evanescent,ettabib2024waveguide}, supported by the maturity of the Si$_3$N$_4$ fabrication platform. Tantalum pentoxide (Ta$_2$O$_5$) provides similarly high refractive-index contrast and strong evanescent interaction with surface-bound analytes; optimized Ta$_2$O$_5$ slot waveguides have shown higher Raman signal collection than comparable Si$_3$N$_4$ structures \cite{ta2o5WERS}. Titanium dioxide (TiO$_2$) has likewise been investigated as a high-index platform for integrated evanescent Raman spectroscopy \cite{evans2015tio2, evans2016tio2}. Aluminum nitride (AlN) has also been demonstrated as a platform for waveguide Raman analysis of benzene derivatives \cite{makela2020aln}. 

To further reduce propagation losses, Raman background, and autofluorescence while extending the transmission window down to ultraviolet wavelengths, alternative low-refractive contrast material platform have been explored. UV-written silica (SiO$_2$) waveguides have shown a low measured Raman background under 660- and 785-nm excitation \cite{silica2023raman}. Similarly, aluminum oxide (Al$_2$O$_3$) has been benchmarked against Si$_3$N$_4$, Ta$_2$O$_5$, and TiO$_2$ for waveguide Raman spectroscopy at 785~nm, showing the lowest Raman background but weaker analyte-signal collection due to its lower refractive index \cite{raza2019highindex}. Its broad transparency, relatively higher refractive index than SiO$_2$  and demonstrated low-loss operation at short wavelengths nevertheless make Al$_2$O$_3$ particularly attractive for extending waveguide Raman spectroscopy into the visible and ultraviolet regions \cite{dullo2025alumina,gok2025evaporated}.

Much of waveguide-coupled Raman spectroscopy has focused on near-infrared excitation, particularly around 785~nm. Moving towards visible and ultraviolet wavelengths offers several potential advantages. The Raman scattering efficiency increases strongly at shorter excitation wavelengths, while ultraviolet excitation can provide access to resonance-enhanced Raman processes for suitable analytes\cite{asher1988uv, efremov2008achievements, fan2010uv}. In integrated photonic platforms, shorter wavelengths can also support reduced optical mode dimensions and higher local optical intensities for appropriately designed waveguides. Together, these benefits can increase the strength of analyte-specific Raman signals. However, operation at shorter wavelengths places substantially greater demands on the waveguide material. Propagation losses and material absorption generally become critical, while fluorescence and device-generated spectral background can obscure weak analyte Raman bands \cite{dhakal2017background}. Visible-wavelength waveguide Raman spectroscopy has, for example, been demonstrated at 532~nm using Ta$_2$O$_5$, including resonance-enhanced detection of hemoglobin \cite{coucheron2019resonance}.

At shorter wavelengths especially moving from visible to UV, waveguide-material selection therefore involves a trade-off between modal confinement, propagation loss, spectral transparency, and intrinsic Raman and fluorescence backgrounds, with the optimal platform depending strongly on the excitation wavelength and target analyte. Previously reported Al$_2$O$_3$ devices exhibited propagation losses below 0.5~dB/cm and measured autofluorescence approximately 200 times lower than that of Si$_3$N$_4$, both at 405~nm \cite{dullo2025alumina}. Previous studies have also characterized Al$_2$O$_3$ waveguide background spectra at 633 and 785~nm \cite{raza2019highindex}. Here, we extend the characterization to 532 and 325~nm for the ALD-grown platform and examine analyte detectability in a back-collected geometry. Although the relatively low refractive index of Al$_2$O$_3$ provides weaker modal confinement than higher-index photonic platforms, its broad transparency, low propagation loss, and low autofluorescence at short wavelengths make it an attractive candidate for extending waveguide-coupled Raman spectroscopy into the visible and ultraviolet regions.

Here, we investigate Al$_2$O$_3$ waveguides for Raman spectroscopy at visible and ultraviolet wavelengths. We compare the spectrally resolved platform backgrounds of Al$_2$O$_3$ and Si$_3$N$_4$ at 532~nm excitation and characterize the Al$_2$O$_3$ background under 325~nm excitation, including its dependence on waveguide width. Motivated by the low propagation loss previously demonstrated in Al$_2$O$_3$ devices at 405~nm \cite{dullo2025alumina}, we further evaluate whether guided visible and ultraviolet excitation can enable back-collected detection of analyte-specific Raman features above the waveguide background. Using microscope oil as a model analyte, we quantify this detectability through signal-to-background ratio and spectral contrast at both excitation wavelengths.

\section{Materials and Methods}
\subsection{Waveguide platforms and geometry}
In this study, we investigated the waveguide generated Raman background of an atomic-layer-deposited (ALD) Al$_2$O$_3$ platform and compared it with that of a low-pressure chemical-vapour-deposited (LPCVD) Si$_3$N$_4$ reference platform. Details of the LPCVD Si$_3$N$_4$ waveguide fabrication are provided in
Supplementary Section~S1.1. The fabrication approach for the Al$_2$O$_3$ platform is described in \cite{dullo2025alumina}. The devices investigated here had a nominal core thickness of 250~nm, rather than the approximately 140~nm thickness illustrated in that earlier study. The Al$_2$O$_3$ waveguides consisted of this core on a SiO$_2$/Si substrate, whereas the Si$_3$N$_4$ reference waveguides had a nominal core thickness of 145~nm. Raman measurements were performed using 20-mm-long strip waveguides. Al$_2$O$_3$ waveguides with nominal widths of 200, 400, 600, and 800~$\mu$m were investigated, while 600~$\mu$m-wide waveguides were used for the direct Al$_2$O$_3$/Si$_3$N$_4$ comparison at 532~nm. Owing to their large lateral dimensions relative to the excitation wavelength, the waveguides support strongly multimode propagation in the lateral direction.

To account for the dependence of the measured signal on coupling position and modal excitation, the focused excitation spot was translated laterally across the input facet and spectra were recorded at multiple positions distributed over the waveguide width (Supplementary Fig.~S2). Spectra from these positions were subsequently used to determine the mean response and its variation. Because the Al$_2$O$_3$ and Si$_3$N$_4$ structures differ in core thickness, refractive index, and modal field distribution, their comparison is reported as a platform-level spectral-background comparison rather than as a measurement of intrinsic material Raman efficiency.

\subsection{Experimental setup}

\begin{figure}[!htbp]
    \centering
    \includegraphics[width=15cm]{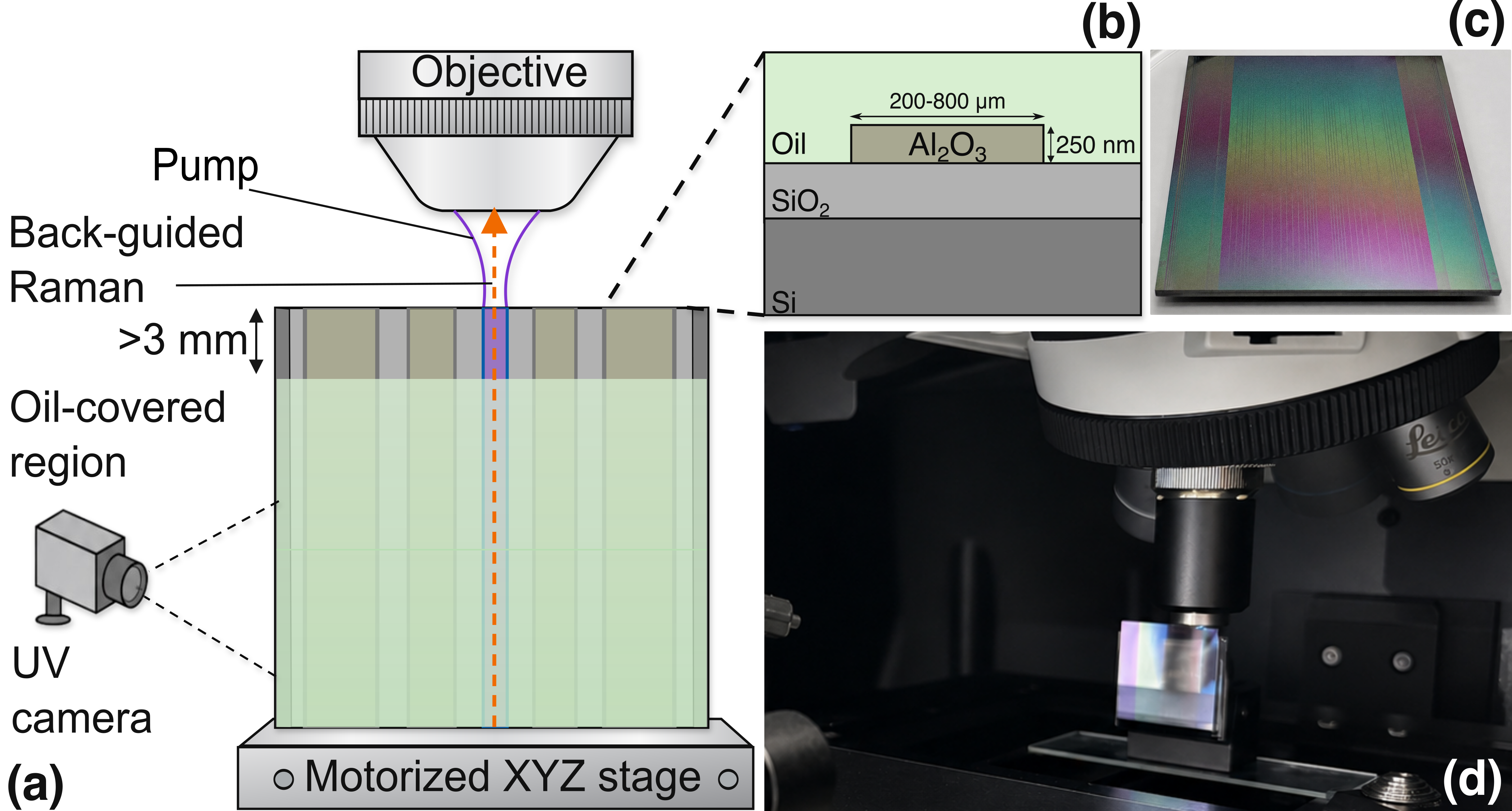}
    \caption{\textbf{Back-collected waveguide Raman configuration.}
(a) Schematic of the measurement geometry. The Al$_2$O$_3$ chip was mounted vertically on a motorized XYZ stage, with the excitation beam coupled into the input facet and back-guided Raman emission collected through the same microscope objective. A UV-sensitive camera viewed the chip along the propagation direction to assist with alignment and coupling. For analyte measurements, immersion oil was deposited on the waveguide surface more than 3~mm from the input facet to prevent direct excitation by the focused input beam. (b) Schematic cross-section of the oil-covered waveguide, showing the 200--800~$\mu$m-wide and nominally 250-nm-thick Al$_2$O$_3$ core on the SiO$_2$/Si substrate. (c) Photograph of the fabricated Al$_2$O$_3$ waveguide chip. (d) Close-up photograph of the vertically mounted chip beneath the microscope objective during coupling and back-collection measurements.}
    \label{fig:experimental_setup}
\end{figure}

Raman measurements were performed using a Renishaw inVia upright microscope modified for waveguide excitation and back-collection. The setup was equipped with a custom sample holder mounted on a motorized three-axis stage, allowing the chip to be positioned vertically with its input facet facing the microscope objective (Fig.~\ref{fig:experimental_setup}). The same objective lens was used for both excitation coupling and collection of the back-propagated Raman signal. A ultraviolet camera imaging along the propagation length was additionally integrated to facilitate optimization of alignment and coupling of the excitation light into the waveguide before each acquisition (Supplementary Fig.~S1). For 325~nm, excitation, a He--Cd Kimmon laser was coupled into  a waveguide using a 40$\times$ ultraviolet objective (Thorlabs LMU-40X-NUV, NA~=~0.47), the Raman signal was analyzed using a 3600-lines/mm grating. For 532-nm excitation, a Leica 50\(\times\) objective (NA = 0.75) and an 1800-lines/mm grating were used.

Spectra were acquired using ten accumulations per measurement. The exposure time and excitation power were selected according to wavelength and sample to avoid detector saturation; the specific acquisition conditions used for each comparison are given in the corresponding Results section. For each waveguide, spectra were recorded at 1--14 coupling positions to account for position-dependent coupling and modal excitation. The number of spectra recorded for each waveguide is shown in Table~\ref{tab:summary_spectral_measurements}. The complete acquisition procedure is described in Supplementary Section~S1.3. For analyte measurements, objective immersion oil (Cargille Laboratories, Custom Laser Liquid, refractive index \(n = 1.504\) at 589.3~nm) was deposited on the waveguide surface. The nearest edge of the oil droplet was positioned more than 3~mm from the input facet to avoid direct illumination of the analyte by the focused coupling beam. The control measurement used to establish this distance is described in Supplementary Section~S1.2. 

Spectral processing depended on the specific analysis. Platform-background comparisons were performed using raw spectra with power--time normalization where indicated. For analyte measurements, unloaded- and oil-loaded-waveguide spectra were scaled or baseline corrected as described in Supplementary Section~S1.4 before subtraction of the waveguide background. Oil-detection performance was quantified from the Raman feature near 1000~cm$^{-1}$ using global and local signal-to-background ratios and a bounded spectral-contrast metric. Definitions of these metrics and the corresponding analysis windows are provided in Supplementary Section~S1.4 and Supplementary Fig.~S3.

\section{Results}
\subsection{Background comparison under 532~nm excitation}

\begin{figure}[!htbp]
    \centering
    \includegraphics[width=8.26cm]{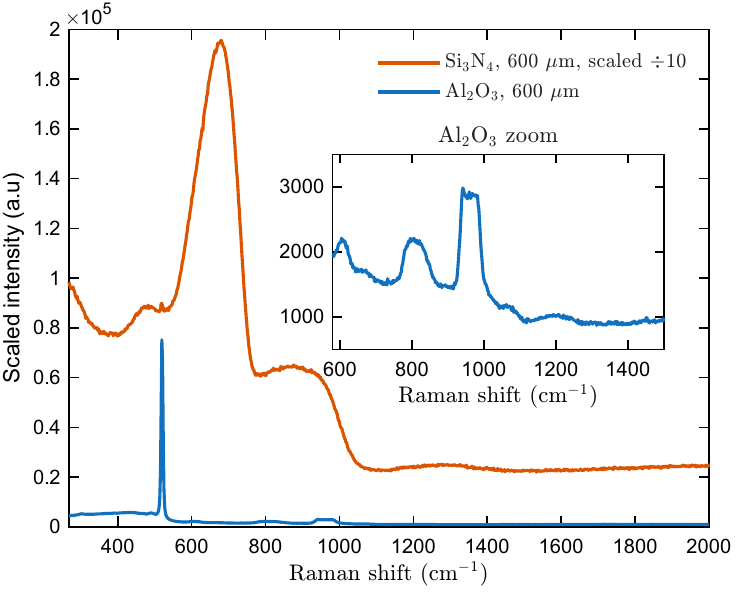}
    \caption{\textbf{Background spectra of Al$_2$O$_3$ and Si$_3$N$_4$ waveguides under 532~nm excitation.}
Spectra were measured from 600~$\mu$m-wide Al$_2$O$_3$ and Si$_3$N$_4$ waveguides with core thicknesses of 250 and 145~nm, respectively. The Si$_3$N$_4$ spectrum was scaled to the Al$_2$O$_3$ power–time product and additionally divided by 10 for display. The inset shows an enlarged view of the Al$_2$O$_3$ spectrum.}
    \label{fig:bg532}
\end{figure}

We first compared the waveguide-generated Raman backgrounds of Al$_2$O$_3$ and Si$_3$N$_4$ under 532~nm excitation (Fig.~\ref{fig:bg532}). The Al$_2$O$_3$ spectrum was acquired at an excitation power of 4.34~mW with a 5~s exposure, whereas the Si$_3$N$_4$ spectrum was acquired at 0.24~mW with a 3~s exposure to avoid detector saturation. Both measurements used 10 accumulations and identical collection settings. Assuming a linear dependence on excitation power and integration time, the Si$_3$N$_4$ spectrum was multiplied by a factor of 30.14, calculated as $(4.34~\mathrm{mW}\times5~\mathrm{s})/(0.24~\mathrm{mW}\times3~\mathrm{s})$, to represent the power--time product used for the Al$_2$O$_3$ measurement. For visual clarity, the scaled Si$_3$N$_4$ spectrum shown in Fig.~\ref{fig:bg532} was additionally divided by 10. This display factor was not applied in the quantitative comparison.

The Si$_3$N$_4$ waveguide exhibits a substantially stronger and more structured background despite the lower excitation power. In particular, a broad band extends from approximately 530 to 770~cm$^{-1}$ with a maximum near 680~cm$^{-1}$, accompanied by an elevated background across much of the fingerprint region. Previous visible-wavelength WERS studies have likewise reported substantial background in Si$_3$N$_4$ waveguides \cite{tyndall2024visible}. The pronounced peak at 520~cm$^{-1}$ and features at 950--1000~cm$^{-1}$ are consistent with crystalline-Si contributions; additional Si peaks near 520 and 960~cm$^{-1}$ have been reported in 532-nm measurements of SiN waveguides \cite{dhakal2017background}. In the present geometry, a possible pathway is a small fraction of the pump reaching the Si substrate near the input facet, although this pathway was not independently established. The weaker features near 600 and 800~cm$^{-1}$ may include contributions from the SiO$_2$ buried oxide or the Al$_2$O$_3$ core. Corresponding substrate, cladding, and core reference measurements would be required to distinguish these contributions.

For quantitative comparison, the raw spectra were integrated over 530-1800~cm$^{-1}$. The ratio of the integrated raw background, Si$_3$N$_4$/Al$_2$O$_3$ was 12.7. After normalization to the respective incident excitation power--time products, this ratio increased to approximately 383, corresponding to a more than two-orders-of-magnitude difference in the normalized measured background. This normalization assumes approximately linear scaling with excitation power and exposure time and does not account for differences in coupling efficiency or modal overlap. The Si$_3$N$_4$ background contains not only Raman scattering but also broadband photoluminescence. Linear power dependence of photoluminescence has been reported for Si$_3$N$_4$ waveguides \cite{jayakumar2025chip}, but this does not independently establish linearity for the present devices and acquisition conditions. The normalized ratio therefore provides an empirical platform-level comparison relevant to waveguide enhanced Raman sensing, rather than a direct measure of intrinsic material Raman efficiency. 

\subsection{\texorpdfstring{Al$_2$O$_3$ waveguide background under 325 nm excitation}{Al2O3 waveguide background under 325 nm excitation}}

\begin{figure}[!htbp]
    \centering
    \includegraphics[width=14.86cm]{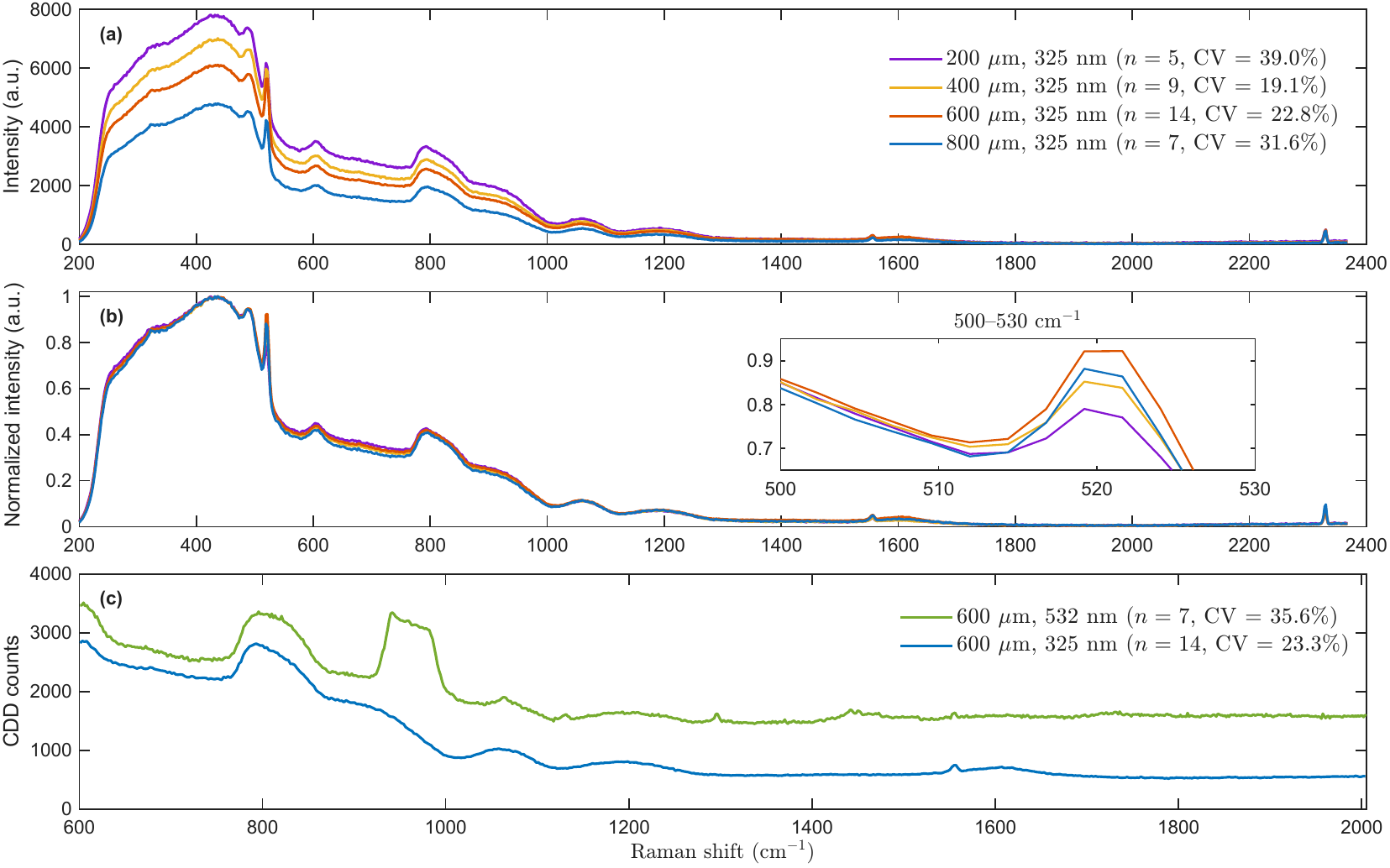}
    \caption{\textbf{Al$_2$O$_3$ waveguide background under ultraviolet and visible excitation.}
    (a) ALS-corrected, non-normalized mean background spectra from waveguides with widths of 200, 400, 600, and 800~$\mu$m under 325~nm excitation. The number of measurements, \(n\), and coefficient of variation (CV = standard deviation divided by the mean of the integrated raw spectral area) are given in the legend.
    (b) The same spectra normalized to their respective maxima. The inset enlarges the 500--530~cm$^{-1}$ region.
    (c) Uncorrected mean background spectra from the 600~$\mu$m-wide waveguide under 325- and 532-nm excitation, shown for qualitative comparison of spectral shape only; acquisition conditions differed between the two wavelengths.}
    \label{fig:width325}
\end{figure}

We next characterized the Raman background of Al$_2$O$_3$ waveguides under ultraviolet excitation. Si$_3$N$_4$ was not characterized at this wavelength because of its limited transmission in the UV. As shown in Fig.~\ref{fig:width325}, the Al$_2$O$_3$ waveguides exhibited a broad background concentrated mainly below approximately 1000~cm$^{-1}$, with weak residual intensity at higher Raman shifts apart from narrow atmospheric features.

We then examined the dependence of this background on waveguide width using 200, 400, 600, and 800~$\mu$m-wide Al$_2$O$_3$ waveguides. Increasing the width by a factor of four did not produce a corresponding increase in background intensity. Instead, the baseline-corrected mean background decreased progressively with increasing waveguide width (Fig.~\ref{fig:width325}(a)). One possible explanation is the different excitation and intensity distribution of the guided modes: wider waveguides support a larger number of modes, allowing the coupled power to be distributed among more modes and potentially reducing the intensity carried by individual modes. The coefficients of variation of 19--39\% (calculated as standard deviation divided by the mean of the integrated raw spectral area) further indicate sensitivity to the coupling conditions. Since the propagated power and modal distribution were not measured independently, however, the observed width dependence should not be interpreted as a universal geometrical scaling law.

After normalization to the respective maxima, the spectra overlap closely over most of the measured range (Fig.~\ref{fig:width325}(b)). Thus, the spectral shape remains broadly similar across the investigated waveguide widths, despite differences in the measured background magnitude. The small differences in the 570-900~cm$^{-1}$ region do not indicate the emergence of a new width-dependent background. The bands near 600 and 800~cm$^{-1}$ are consistent with the SiO$_2$ buried-oxide layer, whereas the line near 520~cm$^{-1}$ is attributed to the crystalline-Si substrate. The narrow features near 1555 and 2330~cm$^{-1}$ arise from atmospheric O$_2$ and N$_2$, respectively.

Figure~\ref{fig:width325}(c) qualitatively compares the uncorrected background from the same 600~$\mu$m-wide waveguide under 325- and 532-nm excitation. Because the measurements were performed under different parameters, the absolute background intensities cannot be directly compared. Nevertheless, the substantially different spectral profile demonstrates that the measured platform background depends strongly on excitation wavelength. Consequently, the suitability of the platform for Raman sensing should be assessed based on the signal-to-background ratio obtained with the target analyte, as discussed in the following section.

Finally, we investigated the contribution of non-coupled light at the chip facet to the measured background by comparing spectra acquired from the same waveguide under optimized coupled and decoupled conditions, and also measured objective background signal (Supplementary Fig.~S4). Coupling produced a pronounced increase in the measured background intensity, whereas several principal spectral features remained detectable under decoupled conditions. These measurements indicate that the observed background contains both coupling-dependent waveguide contributions and parasitic contributions associated with direct illumination of the input-facet region.

\subsection{Back-collected Raman detection of microscope oil}
\begin{figure}[!htbp]
    \centering
    \includegraphics[width=\columnwidth]{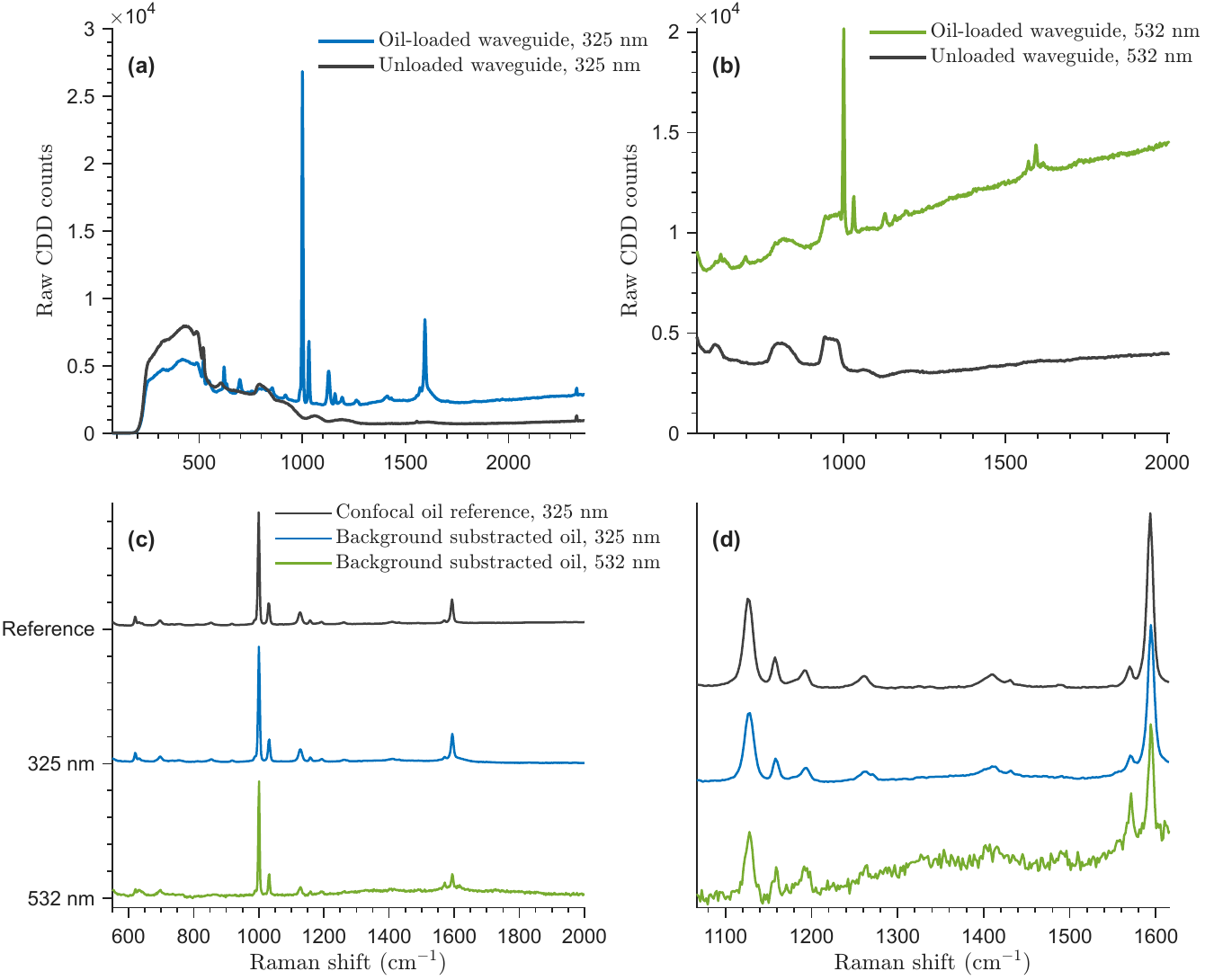}
    \caption{\textbf{Back-collected Raman detection of microscope oil.} Measurements were obtained from a 200~$\mu$m-wide Al$_2$O$_3$ waveguide under 325- and 532-nm excitation.\\
    (a) Raw mean oil-loaded and unloaded-waveguide spectra under 325~nm excitation.\\
    (b) Raw mean oil-loaded and unloaded-waveguide spectra under 532~nm excitation.\\
    (c) Background-subtracted oil spectra obtained under 325 and 532~nm excitation, compared with a confocal reference spectrum of the same oil acquired under 325~nm excitation. The spectra are vertically offset for clarity.\\
    (d) Magnified view of the 1090--1620~cm$^{-1}$ region in panel (c).}
    \label{fig:oil}
\end{figure}

To assess whether analyte-specific Raman features could be recovered above the waveguide background, microscope oil was deposited on Al$_2$O$_3$ waveguides. Representative spectra from a 200~$\mu$m-wide waveguide are shown in Fig.~\ref{fig:oil}, while measurements across the investigated waveguide widths are included in the quantitative comparison in Table~\ref{tab:summary_spectral_measurements}. Raman spectra were acquired in the back-collection geometry under both 325- and 532-nm excitation (see Materials and Methods and Supplementary Section~S1.2). 

For measurements at 325~nm, the unloaded waveguide background was recorded before oil deposition. The oil-loaded and unloaded spectra were acquired using identical excitation power, exposure time, number of accumulations, and collection settings. To enable accurate background subtraction, the oil-loaded and unloaded spectra were normalized such that their common broad background maximum near 441 cm$^{-1}$ — a region without resolved oil-specific Raman features—had the same intensity. This scaling ensures that the waveguide-background contribution underlying the oil-loaded measurement is matched in magnitude to the unloaded-waveguide reference before subtraction. As shown in Fig.~\ref{fig:oil}(a), oil deposition introduced distinct Raman features on top of the waveguide background. The most prominent additional bands occur near 1000~cm$^{-1}$ and 1030~cm$^{-1}$, with additional features between 1100--1200~cm$^{-1}$ and near 1600~cm$^{-1}$. Subtracting the scaled unloaded-waveguide spectrum from the oil-loaded spectrum yielded the analyte-associated contribution shown by the blue trace in Fig.~\ref{fig:oil}(c). The positions of its principal bands agree with those in the confocal oil reference by the black trace in Fig.~\ref{fig:oil}(c), supporting assignment of the additional signal to the deposited microscope oil.

The same 200~$\mu$m-wide waveguide was subsequently measured under 532~nm excitation (Fig.~\ref{fig:oil}(b)). A coupling-sensitive broadband slope was occasionally observed at 325~nm but was consistently present at 532~nm. Since its origin could not be established, the oil-loaded and unloaded spectra for 532~nm excitation (Fig.~\ref{fig:oil}(b)) were independently baseline corrected and scaled to their respective intensities near 944~cm$^{-1}$ before subtraction. The resulting difference spectrum is shown by the green trace in Fig.~\ref{fig:oil}(c).

The background-subtracted spectra obtained at both excitation wavelengths reproduce the characteristic Raman fingerprint of the microscope oil, with a dominant band near 1000~cm$^{-1}$ and additional features whose positions correspond to bands in the confocal reference. The secondary bands near 1030~cm$^{-1}$, in the 1100--1200~cm$^{-1}$ region, and near 1600~cm$^{-1}$ are more clearly resolved at 325~nm. The corresponding features are also present in the 532~nm spectrum, although with higher noise and different relative band intensities, particularly in the expanded 1090--1620~cm$^{-1}$ region shown in Fig.~\ref{fig:oil}(d). The stronger analyte feature at 325~nm is consistent with the stronger Raman scattering expected at shorter excitation wavelengths. Differences in relative band intensity may additionally reflect excitation and collection efficiencies, modal excitation, detector response, and the effects of baseline correction and scaling.

The confocal reference exhibits the same principal oil fingerprint, although its relative band intensities and baseline differ because the confocal and waveguide configurations use different excitation and collection geometries. The agreement between the waveguide-derived and confocal spectra demonstrates that the characteristic Raman signature of the deposited oil can be recovered through the waveguide-mediated excitation and collection geometry under both 325~nm and 532~nm excitations.

Peak-based metrics for the oil band near 1000~cm$^{-1}$ are summarized in Table~\ref{tab:summary_spectral_measurements}. The local signal-to-background ratio (SBR) compares the raw peak intensity with the local baseline estimated from the endpoints of the peak window, and therefore quantifies how clearly the Raman band emerges from the underlying local spectral background. The global SBR compares the raw peak intensity with the corresponding unloaded-waveguide background at the same peak position. This metric evaluates the contrast between the oil-coated waveguide signal and the external/background signal, and is therefore useful for assessing the overall experimental detectability of the oil signal. Finally, the contrast metric, defined from the peak height relative to the sum of the peak and baseline intensities, provides a 0-to-1 bounded, normalized measure of peak visibility that is less directly dependent on absolute signal magnitude.

\begin{table}[!htbp]
    \centering

    \caption{1000~cm$^{-1}$ peak-based metrics for microscope-oil detection using
    Al$_2$O$_3$ waveguides under 325- and 532-nm excitation.
    $N_{\mathrm{bg}}$ and $N_{\mathrm{oil}}$ denote the numbers of
    unloaded- and oil-loaded-waveguide spectra, respectively.}

    \label{tab:summary_spectral_measurements}

    \begin{tabular}{
        S[
            table-format=3.0,
            round-mode=places,
            round-precision=0,
            round-pad=false
        ]
        S[
            table-format=3.0,
            round-mode=places,
            round-precision=0,
            round-pad=false
        ]
        S[
            table-format=2.0,
            round-mode=places,
            round-precision=0,
            round-pad=false
        ]
        S[
            table-format=2.0,
            round-mode=places,
            round-precision=0,
            round-pad=false
        ]
        S[
            table-format=4.1,
            round-mode=places,
            round-precision=1,
            round-pad=true
        ]
        S[
            table-format=2.2,
            round-mode=places,
            round-precision=1,
            round-pad=true
        ]
        S[
            table-format=1.2,
            round-mode=places,
            round-precision=2,
            round-pad=true
        ]
    }

        \toprule

        {\textbf{Excitation}} &
        {\textbf{Width}} &
        {\textbf{$N_{\mathrm{bg}}$}} &
        {\textbf{$N_{\mathrm{oil}}$}} &
        {\textbf{SBR global}} &
        {\textbf{SBR local}} &
        {\textbf{Contrast}} \\

        {\textbf{(nm)}} &
        {\textbf{($\mu$m)}} &
        {} &
        {} &
        {} &
        {} &
        {} \\

        \midrule

        325 & 200 & 5  & 6  
            & 23.26  & 9.1133
            & 0.80224 \\

        {}  & 400 & 9  & 5  
            & 12.219 & 7.6253
            & 0.76812 \\

        {}  & 600 & 14 & 12 
            & 12.791 & 6.4280
            & 0.73075 \\

        {}  & 800 & 7  & 4  
            & 9.2417 & 6.0509
            & 0.71635 \\

        \midrule

        532 & 200 & 5  & 5  
            & 5.6549 & 1.92612
            & 0.31650 \\

        {}  & 400 & 4  & 6  
            & 5.7427 & 1.25678
            & 0.11378 \\

        {}  & 600 & 5  & 6  
            & 2.6415 & 1.21666
            & 0.09774 \\

        \bottomrule

    \end{tabular}
\end{table}

Across all measured waveguides widths, both the SBR and contrast were higher under 325~nm excitation than under 532~nm excitation. In particular, local SBR values of 6.05--9.11 and contrast values of 0.72--0.80 under 325~nm excitation demonstrate that the oil Raman bands remain clearly resolved above the local waveguide background. These quantitative metrics provide direct experimental evidence that the low-background Al$_2$O$_3$ platform enables recovery of analyte Raman signatures under ultraviolet wavelength excitation.

\subsection{Detection of trace PET-consistent contamination}
After the oil measurements, the waveguides were cleaned with isopropyl alcohol and tested for  reproducibility of the background signal. Here, weak narrow bands appeared at approximately 635, 860, 1290, 1615, and 1727~cm$^{-1}$. These features were absent from the freshly fabricated waveguides before oil deposition and cleaning, as shown by the corresponding before-and-after spectra in Supplementary Fig.~S5. The five bands do not match the expected Raman responses of the Al$_2$O$_3$ core, SiO$_2$ buried oxide, or Si substrate, but their positions closely correspond to the characteristic Raman fingerprint of polyethylene terephthalate (PET) \cite{penalver2023raman,xu2021temperature}. Their simultaneous appearance therefore supports assignment to a trace PET-consistent residue. Its appearance after the oil measurements and cleaning suggests introduction during deposition, cleaning, or subsequent handling, but the spectra do not establish the precise route. No corresponding PET bands were observed in confocal measurements of the oil or waveguide top surface under the conditions used. The observation additionally illustrates the sensitivity of the waveguide-coupled geometry to trace surface contamination that was not detected in the corresponding confocal measurements, consistent with the extended interaction length provided by waveguide-mediated excitation and collection.

\section{Discussion and Conclusion}

The results demonstrate waveguide-mediated excitation and back-collection of analyte Raman signals using highly multimode Al$_2$O$_3$ waveguides at both 325 and 532~nm. In contrast to conventional single-mode WERS platforms, the present 200--800~$\mu$m-wide waveguides support a large number of guided modes, allowing excitation of the oil-covered region and recapture of a fraction of the generated Raman emission into backward-propagating modes. The measurements further show that the detected background is determined not only by guided propagation, but also by contributions from the input-facet region, buried oxide, optical collection path, and sample history.

At 532~nm, Al$_2$O$_3$ exhibited a substantially weaker and less structured measured background than Si$_3$N$_4$, supporting its suitability as a low-background platform for visible Raman sensing. Under ultraviolet excitation, analyte-specific Raman bands were clearly recovered above the Al$_2$O$_3$ waveguide background. For the waveguide widths investigated at both wavelengths, the local SBR was consistently higher at 325~nm than at 532~nm. Under ultraviolet excitation, the local SBR ranged from 6 to 9, with corresponding contrast of 0.7--0.8. Because the two wavelengths were measured using different optical configurations, these results do not establish the effect of excitation wavelength alone on Raman detection efficiency. Nevertheless, these results demonstrate that Al$_2$O$_3$ waveguides enable ultraviolet Raman detection of analytes, with characteristic bands clearly distinguishable from the waveguide background.

The measured intensity and its width dependence reflect the combined effects of coupling, modal excitation and collection, and  should therefore be interpreted as a response of the complete coupling--waveguide--collection system. Suspending the Al$_2$O$_3$ waveguide would remove the underlying SiO$_2$ and Si from the optically active region and could therefore reduce substrate-derived Raman background. Centimetre-scale suspended waveguides fabricated from ALD-grown Al$_2$O$_3$ have already been demonstrated \cite{gehl2021characterization}, supporting the feasibility of this approach, although the resulting reduction in Raman background remains to be quantified. Overall, the combination of low measured background, short-wavelength transparency, and successful back-collected Raman detection establishes Al$_2$O$_3$ as a promising platform for extending waveguide Raman spectroscopy into the ultraviolet.

\section*{Disclosures}
The authors declare no conflicts of interest.

\section*{Acknowledgments}
RZ thanks Kåre Edvardsen for providing the UV-sensitive camera used to monitor waveguide coupling. R.Z. acknowledges support from the REMARCO project (“Remediation, Management, Monitoring and Cooperation addressing North Sea UXO”), co-funded by the European Union through the European Regional Development Fund under the Interreg North Sea Program 2021–2027. BSA acknowledges the funding from the Research Council of Norway (Project \#352764) and EU Chip JU. E.V.B acknowledges Erasmus+ exchange program.

\noindent\textbf{Supplementary Information:}
Si$_3$N$_4$ waveguide fabrication; camera-assisted coupling optimization and analyte-placement control; coupling-position sampling; spectral processing and baseline-correction parameters; peak-metric definitions; coupled, decoupled, and objective-background spectra; and PET-consistent residue spectra (PDF). The supplementary information accompanies this preprint.

\section*{Data availability}
Data underlying the results presented in this paper may be obtained from the authors upon reasonable request.

\small
\setlength{\bibsep}{3pt plus 1pt minus 1pt}
\bibliographystyle{unsrtnat}
\bibliography{references}
\end{document}


\maketitle
\thispagestyle{empty}

\begin{center}
    \small
    $^{1}$Department of Physics and Technology, UiT The Arctic University of
    Norway, Klokkargårdsbakken 35, NO-9037 Tromsø, Norway\\[0.5em]
    $^{2}$SINTEF Digital Dept. of Microsystems and Nanotechnology, 
    Gaustadallen 23C, 0373 Oslo, Norway\\[0.5em]

    $^{*}$\href{mailto:roman.zakoldaev@uit.no}
    {roman.zakoldaev@uit.no}\\
    $^{**}$\href{mailto:Firehun.t.dullo@sintef.no}
    {Firehun.t.dullo@sintef.no}\\
    $^{***}$\href{mailto:balpreet.singh.ahluwalia@uit.no}
    {balpreet.singh.ahluwalia@uit.no}\\[0.5em]
\end{center}

\tableofcontents

\section{Experimental setup and spectral processing}

\subsection{Fabrication of the LPCVD Si$_3$N$_4$ waveguides}

The fabrication of the LPCVD $\mathrm{Si_3N_4}$ waveguides started with a 6-inch silicon wafer coated with a 2.5-$\mu$m-thick thermally grown $\mathrm{SiO_2}$ layer at 1500~$^\circ$C, which served as a buffer layer to suppress optical leakage into the underlying silicon substrate. Subsequently, a 145-nm-thick $\mathrm{Si_3N_4}$ layer was deposited by low-pressure chemical vapor deposition (LPCVD) at 750~$^\circ$C. The waveguide structures were then defined using photolithography, followed by dry etching of the $\mathrm{Si_3N_4}$ layer to form two-dimensional (2D) channel waveguides. After patterning, the $\mathrm{Si_3N_4}$ waveguides were protected with a $\mathrm{SiO_2}$ cladding layer deposited by plasma-enhanced chemical vapor deposition (PECVD) at low temperature. Finally, selected regions of the $\mathrm{SiO_2}$ cladding were opened to expose the underlying waveguides for optical imaging. This was achieved using a combination of dry and wet etching processes, providing optical access to the waveguides while maintaining the protective oxide cladding over the remaining chip area.

\subsection{Chip mounting, coupling optimization, and analyte placement}

The chip was mounted vertically in a custom holder attached to the motorized three-axis translation stage of the Renishaw inVia Raman microscope. The input waveguide facet faced the microscope objective, which was used both to couple the excitation into the waveguide and to collect Raman emission returning to the input facet.

Coupling was optimized by monitoring propagation scattering along the waveguide with an external camera based on a Raspberry Pi camera module modified for UV visualization. The camera was positioned to capture the longest accessible section of the waveguide while excluding the saturated region near the input facet. The camera region of interest extended over more than 1.3 cm of the waveguide downstream from the coupling region. The mean pixel intensity within this region was used as a proxy for guided-light propagation, and the coupling position was adjusted to maximize this value.

At each nominal position across the input facet, the motorized stage coordinates were adjusted while the camera signal was monitored. The coupling condition was optimized until the mean pixel intensity within the selected region, outlined in green in the figure, reached its highest reproducibly attainable value. The same optimization criterion was applied at each successive coupling position. Representative images of slightly decoupled and optimized coupling conditions for a \(600~\si{\micro\metre}\)-wide waveguide under 325-nm excitation are shown in Supplementary Fig.~\ref{fig:camera-coupling}.

\begin{figure}[ht]
    \centering
    \includegraphics[width=14.5cm]{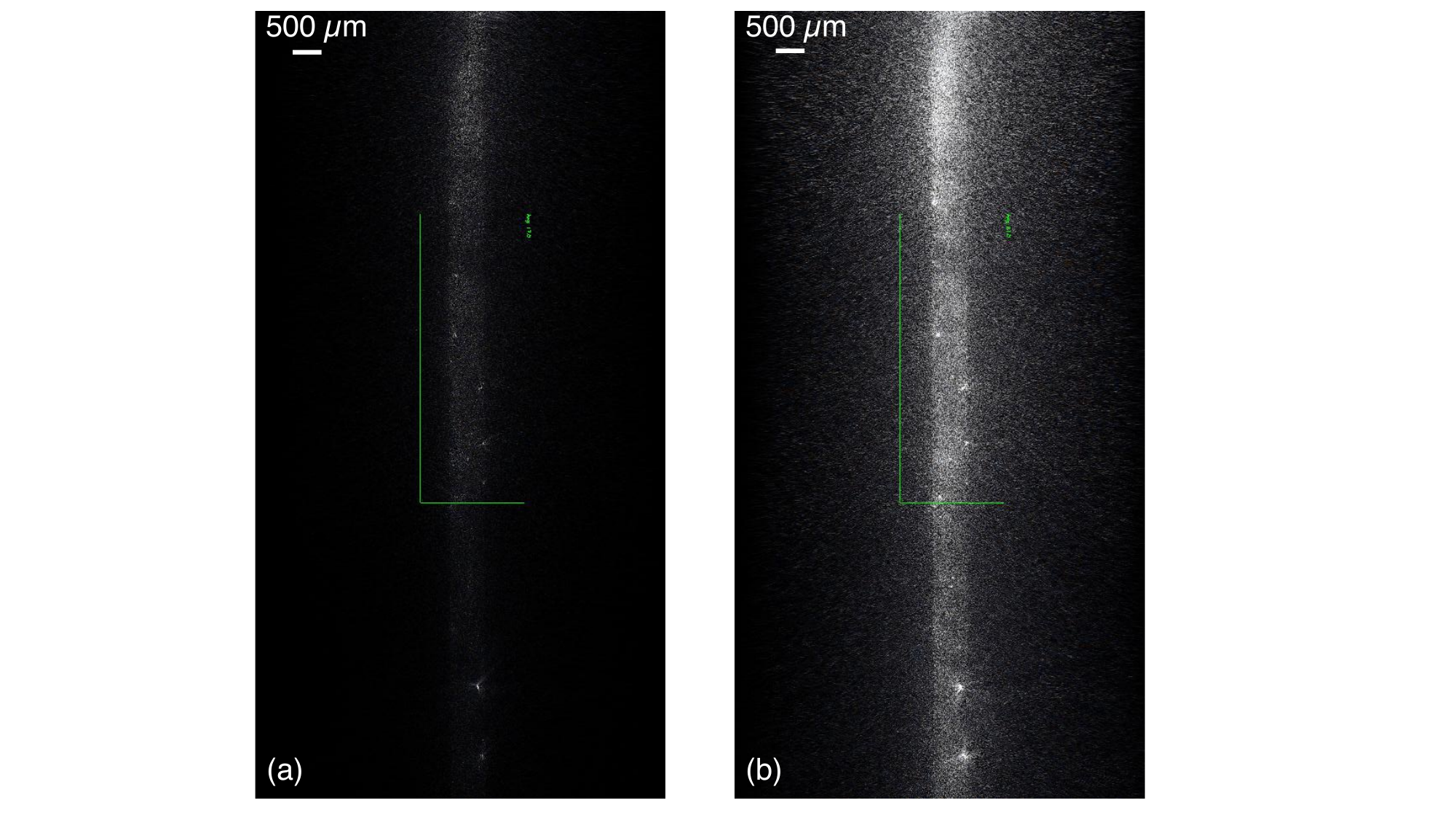}
    \caption{\textbf{Camera-assisted optimization of 325-nm coupling into a \(600~\si{\micro\metre}\)-wide waveguide.}
    Representative camera images acquired under (a) a slightly decoupled condition and (b) the optimized coupling condition. The region used to evaluate propagation scattering is outlined in green. Both images were acquired and displayed using identical camera settings. The scale bar represents \(500~\si{\micro\metre}\).}
    \label{fig:camera-coupling}
\end{figure}

For analyte measurements, oil was deposited on the top surface of the waveguide more than 3 mm from the input facet. This distance was sufficient to prevent direct illumination of the oil by the focused coupling beam. As a control, we first determined the distance over which direct excitation by the focused beam could produce detectable Raman signal. Using the same objective in a conventional microscope configuration, with the spectrometer and collection path otherwise unchanged, the oil was positioned at increasing distances from the focal point. No oil Raman bands were detected when the oil was more than 2 mm from the focus. This experimentally determined distance was therefore used to exclude direct excitation of the oil in the waveguide measurements. The observed oil Raman signal consequently supports propagation of the excitation along the waveguide to the oil-covered region, followed by return of a fraction of the generated Raman emission toward the input facet.

Oil deposition changes the refractive index of the upper cladding. The oil-loaded and oil-free waveguides are therefore not optically identical, and subtraction of the corresponding spectra should be regarded as an empirical procedure for recovering analyte-associated spectral features.

\subsection{Coupling positions and repeated acquisitions}

Each recorded spectrum comprised ten accumulations summed by the acquisition software. For each waveguide, spectra were acquired at \(n_{\mathrm{sp}}=1\)--14 independently optimized coupling positions distributed across the input facet. The nominal separation between successive coupling positions was approx. \(50~\si{\micro\metre}\) or \(100~\si{\micro\metre}\). A nominal separation of \(50~\si{\micro\metre}\) was used for the narrower waveguides to obtain a larger number of measurements across the  width. Because both stage coordinates were adjusted during coupling optimization, the final positions did not necessarily follow the nominal separation exactly. The translation-stage coordinates were expressed relative to the first successful coupling position for each waveguide:
\begin{equation}
    \Delta\mathbf{r}_{i}
    =
    \mathbf{r}_{i}-\mathbf{r}_{0},
    \label{eq:relative-coupling-position}
\end{equation}
where \(\mathbf{r}_{i}\) is the optimized stage position for measurement \(i\), and \(\mathbf{r}_{0}\) is the first successful coupling position. Representative relative positions for the 600-\(\si{\micro\metre}\)-wide waveguide are shown in Supplementary Fig.~\ref{fig:coupling-positions}. The correlated changes in the two stage coordinates are consistent with a small misalignment between the chip orientation and the nominal stage-translation direction.

\begin{figure}[!htbp]
    \centering
    \includegraphics[width=\linewidth]{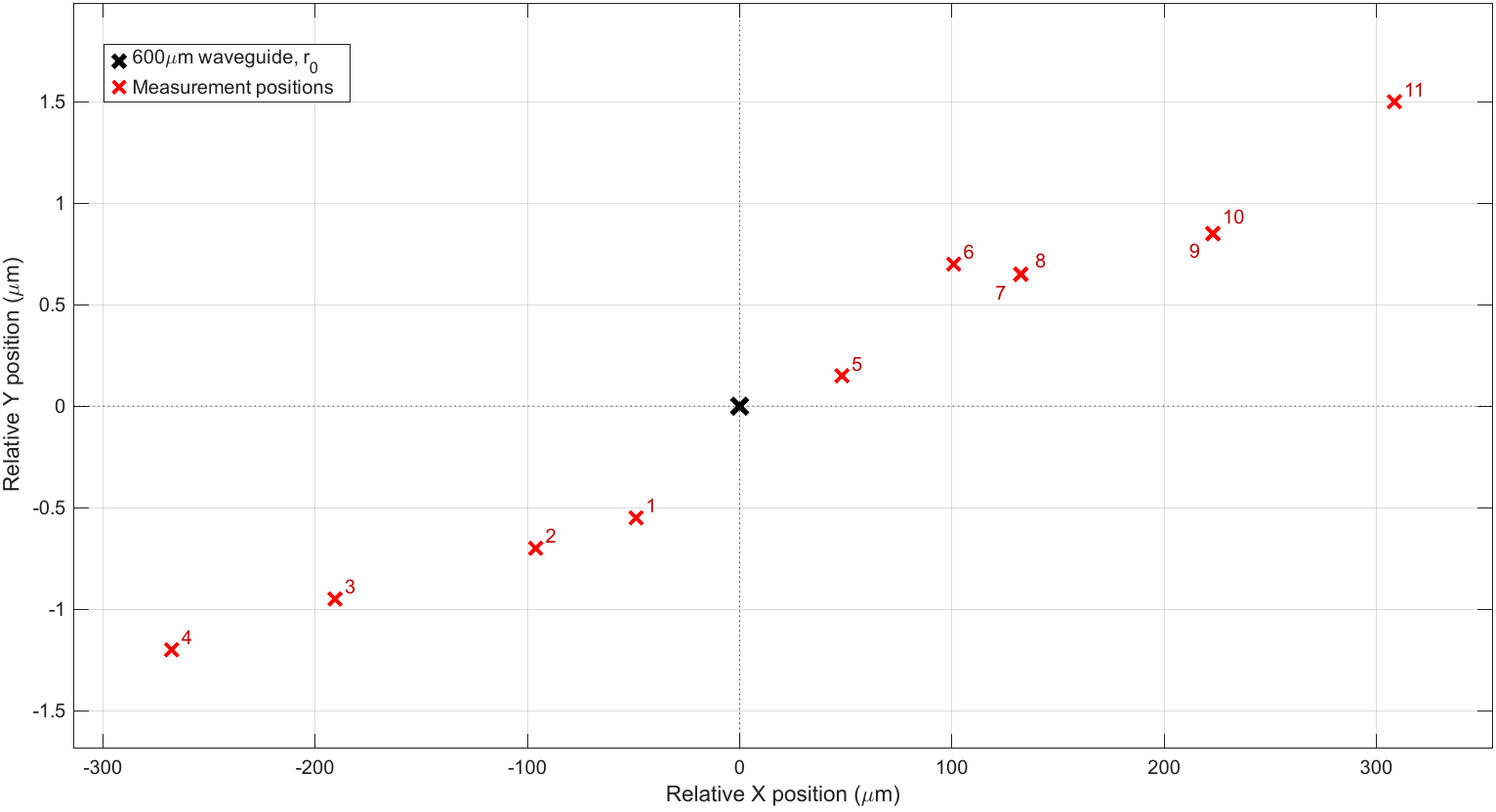}
    \caption{\textbf{Relative coupling positions used for Raman acquisition from the 600-\(\si{\micro\metre}\)-wide oil-loaded waveguide for 325 nm excitation.}
    Coordinates are shown relative to the first successful coupling position (\(\mathbf{r}_{0}\)). Each point represents an independently optimized position at which one spectrum was recorded (total of $N_{oil}=12$ measurement).}
    \label{fig:coupling-positions}
\end{figure}

\subsection{Spectral processing}

Spectra were exported from Renishaw WiRE~5.5 and processed in MATLAB R2026a. No smoothing was applied. Spectra acquired at independently optimized coupling positions were averaged pointwise for each waveguide and measurement condition.

Spectral data were processed as follows. In Fig.~2 of the main manuscript, raw mean spectra were compared without baseline correction or normalization. The Si$_3$N$_4$ spectrum was corrected for the difference in excitation power and exposure time and was additionally divided by 10 for display. For Fig.~3(a,b) of the main manuscript, the mean background spectra were independently corrected by asymmetric least-squares baseline subtraction using an asymmetry parameter of \(p=0.001\), a smoothness parameter of \(\lambda=10^{8}\), and 100 iterations. The spectra displayed in normalized form were divided by their respective maxima. For Fig.~4(c,d) of the main manuscript, the mean oil-loaded and oil-free background spectra were independently baseline-corrected and normalized at 441~cm$^{-1}$ for 325-nm excitation and 944~cm$^{-1}$ for 532-nm excitation. The resulting background spectrum was subsequently subtracted from the corresponding oil-loaded spectrum. Vertical offsets and any additional scaling applied to the reference spectrum were used only for visualization. The parameters used are detailed in Table~\ref{tab:ALS_parameters}.

\begin{table}[htbp]
\centering
\caption{ALS baseline-correction and normalization parameters used for the different spectral datasets.}
\label{tab:ALS_parameters}

\begin{tabular}{c l c c c c}
\hline
Excitation & Dataset & $\lambda$ & $p$ & Iterations & Reference \\
\hline
325 nm & Oil
& $10^{8}$ & 0.001 & 100 & 441~cm$^{-1}$ \\

325 nm & Background
& $10^{8}$ & 0.001 & 100 & 441~cm$^{-1}$ \\

532 nm & Oil
& $10^{7}$ & 0.001 & 100 & 944~cm$^{-1}$ \\

532 nm & Background
& $10^{5}$ & 0.001 & 100 & 944~cm$^{-1}$ \\

325 nm & Confocal oil reference
& $10^{7}$ & 0.001 & 100 & -- \\
\hline
\end{tabular}

\end{table}

The oil peak was evaluated within 990--1010~cm$^{-1}$. A local linear baseline was constructed between the endpoints of this interval, and the peak position was identified from the maximum intensity. The peak intensity, \(I_{\mathrm{P}}\) and \(I_{\mathrm{Bg}}\) were the intensity of, respectively the raw mean oil spectrum and the raw mean background spectrum at this position, whereas \(I_{\mathrm{b}}\) was the corresponding value of the local baseline of the oil signal. Only global and local signal to background ratio (SBR) and spectral contrast (C) were retained as performance metrics. The metric definitions are given below, and a representative processing workflow is shown in Supplementary Fig.~\ref{fig:metrics}. Waveguide performance was quantified as
\begin{equation}
\mathrm{SBR_{global}}=\frac{I_{\mathrm{P}}}{I_{\mathrm{Bg}}},
\qquad
\mathrm{SBR_{local}}=\frac{I_{\mathrm{P}}}{I_{\mathrm{b}}},
\qquad
C=\frac{I_{\mathrm{P}}-I_{\mathrm{b}}}
        {I_{\mathrm{P}}+I_{\mathrm{b}}}.
\end{equation}

The global ratio quantifies the raw oil-loaded peak intensity relative to the independently measured unloaded-waveguide background. Signal-to-background metrics have also been used in previous WERS studies \cite{dhakal2017background}, although differing signal definitions and measurement conditions limit direct numerical comparison. However, local SBR and Contrast were introduced to measure precisely how good the peak was from the signal itself since its corresponding background signal was not exactly aligned with the oil signal.

\begin{figure}[htbp]
    \centering
    \includegraphics[width=\linewidth]{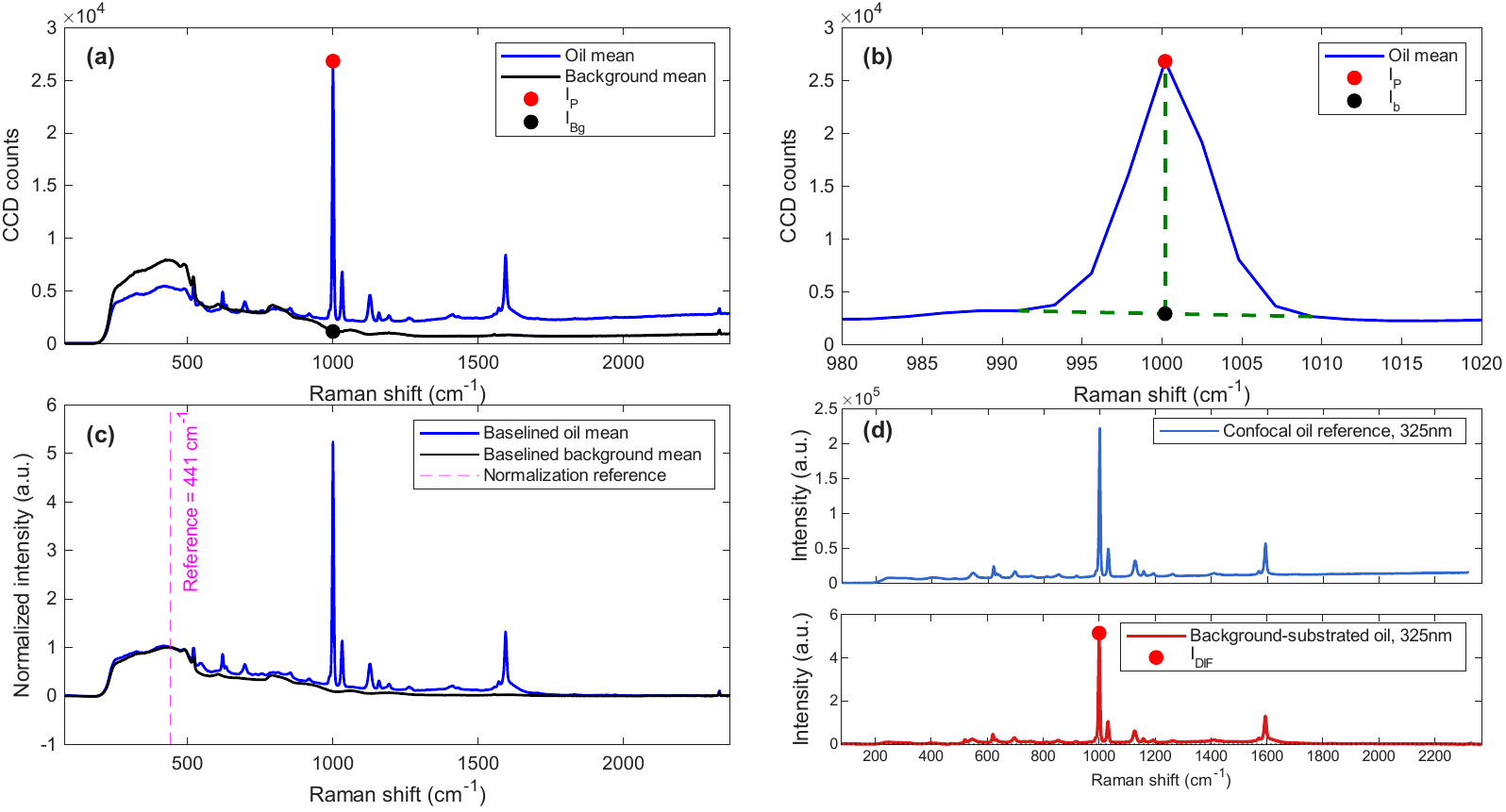}
    \caption{\textbf{Representative spectral processing and peak-metric extraction for the 200-\(\si{\micro\metre}\)-wide waveguide at 325~nm.}
    (a) Raw oil-loaded and oil-free background spectra.
    (b) Construction of the local baseline within the 990--1010~cm\(^{-1}\) interval.
    (c) Normalized and baseline-corrected spectra.
    (d) Background-subtracted oil spectra obtained under 325~nm excitation, compared with a confocal reference spectrum of the same oil acquired under 325~nm excitation, $I_{DIF}$ corresponding to the intensity value at 1000 cm\(^{-1}\) of the background-subtracted oil spectra
    }
    \label{fig:metrics}
\end{figure}

\section{Influence of coupling on the ultraviolet waveguide background}

\begin{figure}[H]
    \centering
    \includegraphics[width=0.7\linewidth]{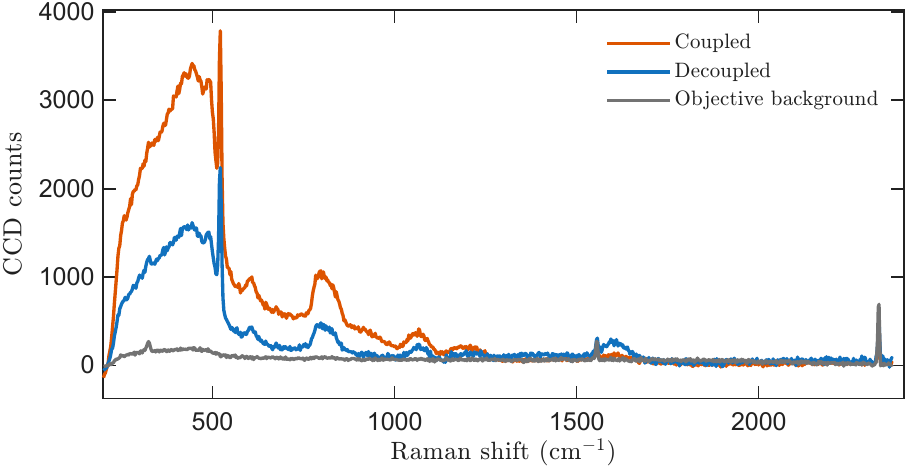}
    \caption{\textbf{Influence of coupling on the ultraviolet Al$_2$O$_3$ waveguide background.}
    Coupled, deliberately decoupled, and objective-background spectra acquired using identical excitation power, exposure time, number of accumulations, and collection settings. The spectra are presented on the same CCD-count scale without normalization or amplitude scaling.}
    \label{fig:coupling-background}
\end{figure}

To examine the influence of light coupling on the measured ultraviolet background, spectra were acquired from the same Al$_2$O$_3$ waveguide under optimized coupled and deliberately decoupled conditions. An independently measured objective background was included for comparison. All spectra were acquired using the same excitation power, exposure time, number of accumulations, and collection settings and are presented on the same CCD-count scale without normalization (Supplementary Fig.~\ref{fig:coupling-background}).

The coupled spectrum exhibits a substantially higher broadband intensity than the decoupled spectrum, particularly below approximately \(1200~\si{\per\centi\metre}\). The objective background is weaker over most of the measured range. Under the conditions employed, the objective alone therefore does not account for the magnitude of the background recorded during waveguide coupling.

Several spectral features persist under decoupled conditions, including the Si line near \(520~\si{\per\centi\metre}\) and atmospheric bands near 1550 and \(2330~\si{\per\centi\metre}\). Their presence shows that these features are not unique to efficient waveguide coupling and can also be detected through uncoupled or scattered excitation of the input facet, substrate, or surrounding device structure.

With the excitation fully decoupled from the waveguide, the substantial remaining signal demonstrates that a major fraction of the measured background originates from direct illumination of the input-facet region and its back-collection through the objective. This contribution may include Raman scattering or fluorescence from the Al$_2$O$_3$ core, buried oxide, Si substrate, and surrounding chip structure. The additional intensity observed under coupled conditions therefore represents a coupling-dependent contribution superimposed on this facet-region background.

\section{Identification of PET-associated Raman bands}

The Al$_2$O$_3$ spectrum presented in Fig.~2 of the main manuscript contains weak additional bands that were absent from spectra acquired from the initially fabricated waveguide chips. These features were first observed after deposition of a \(5~\si{\micro\litre}\) oil droplet and subsequent cleaning with isopropyl alcohol (IPA).

\begin{figure}[H]
    \centering
    \includegraphics[width=0.80\linewidth]{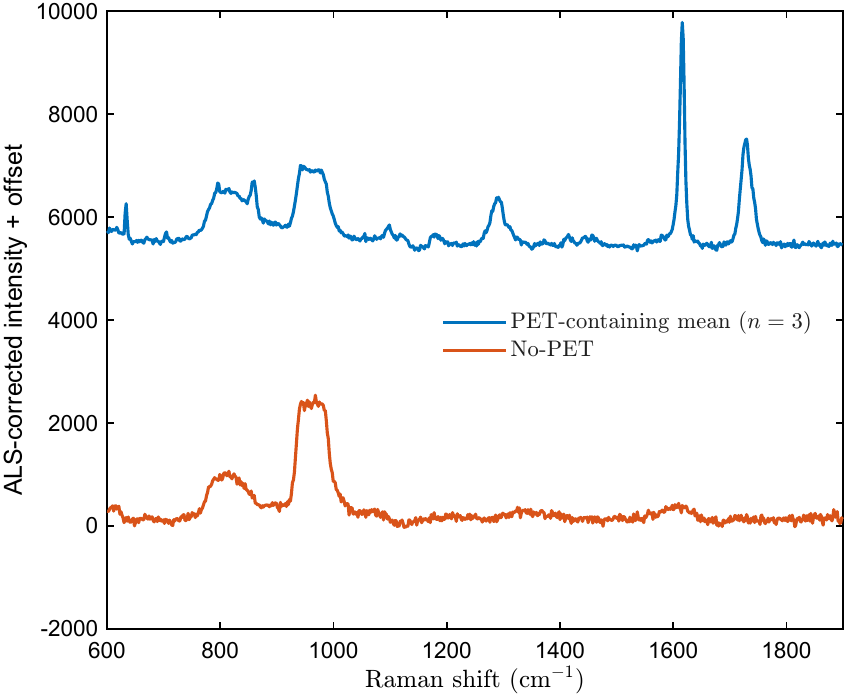}
    \caption{\textbf{PET-consistent Raman bands following the oil measurements.}
    Mean waveguide-coupled spectra from a 600~\si{\micro\metre}-wide Al$_2$O$_3$ waveguide, classified as PET-containing (\(n=3\)) or no-PET. Vertical markers indicate bands near 634, 860, 1291, 1615, and 1727~\si{\per\centi\metre}. The spectra were independently corrected using asymmetric least-squares (ALS) baseline correction and averaged without normalization; the PET-containing mean was vertically offset
    for clarity.}
    \label{fig:pet-control}
\end{figure}

Figure~\ref{fig:pet-control} compares spectra classified according to the presence or absence of these additional features. The simultaneous occurrence of the five marked bands closely matches the characteristic Raman fingerprint of polyethylene terephthalate (PET) \cite{xu2021temperature,penalver2023raman}. The additional signal is therefore assigned to a PET-consistent residue rather than to the intrinsic Raman response of the Al$_2$O$_3$ core, SiO$_2$ buried oxide, or Si substrate.

No corresponding PET bands were detected in conventional confocal measurements under the conditions used in this study. Their detection shows that a low-level PET-consistent residue was detectable in the waveguide-coupled configuration. Together with their absence from the initially fabricated chips, this observation indicates that the bands are not persistent components of the waveguide-material background.

The chronological association with oil deposition and IPA cleaning suggests that the residue was introduced during sample deposition, cleaning, or subsequent handling. Raman spectroscopy does not, however, establish its precise route of introduction. The assignment therefore concerns the chemical identity of the detected residue and does not imply that the oil itself was its source.

\FloatBarrier
\small
\setlength{\bibsep}{3pt plus 1pt minus 1pt}
\bibliographystyle{unsrtnat}
\bibliography{references}